\documentclass{sig-alternate}
\usepackage{booktabs}
\usepackage{xspace}
\usepackage{float}
\usepackage{multirow}
\usepackage{caption}

\usepackage{subfigure}

\usepackage{pgf}

 \usepackage{graphicx}
\usepackage{adjustbox}
 \usepackage{tikz}

\usetikzlibrary{shapes.geometric}

\usetikzlibrary{shapes.multipart}
\usetikzlibrary{chains}

\usepackage{balance}

\usepackage{amsmath}
\usepackage{verbatim}

\usepackage{listings}

\usepackage{color}

\usepackage{calc}

\pgfmathsetseed{1138}

\ifdefined\dontprintsemicolon
\else
\newcommand{\dontprintsemicolon}{\DontPrintSemicolon} 
\fi

\ifdefined\linesnumberedhidden
\else
\newcommand{\linesnumberedhidden}{\LinesNumberedHidden} 
\fi

\definecolor{dkgreen}{rgb}{0,0.6,0}
\definecolor{gray}{rgb}{0.5,0.5,0.5}
\definecolor{mauve}{rgb}{0.58,0,0.82}
 
\def\acknowledgement{This work has been partially supported by the Excellence
 Cluster on Multimodal Computing and Interaction (MMCI) and the  German Research Foundation (DFG) under grant MI 1794/1-1.}

\title{An LSH Index for Computing 
Kendall's Tau \\over Top-k Lists\titlenote{\acknowledgement}}

 \numberofauthors{2}

\author{
\alignauthor
Koninika Pal\\ 
    \affaddr{Saarland University}\\
    \affaddr{Saarbr{\"u}cken, Germany}\\ 
    \email{kpal@mmci.uni-saarland.de} 
\alignauthor
Sebastian Michel\\ 
    \affaddr{Saarland University}\\
    \affaddr{Saarbr{\"u}cken, Germany}\\ 
    \email{smichel@mmci.uni-saarland.de}
}

\conferenceinfo{}{Seventeenth International Workshop on the Web and Databases (WebDB 2014), 
June 22, 2014 - Snowbird, UT, USA.}

\toappear{Copyright is held by the author/owner. \newline Seventeenth International Workshop on the Web and Databases (WebDB 2014), 
June 22, 2014 - Snowbird, UT, USA. }

\global\boilerplate={Copyright is held by the author/owner. }

\floatstyle{plain}
\newfloat{myalgorithm}{thp}{plain}
\floatname{myalgorithm}{Algorithm\xspace}

\def\naive{na\"{\i}ve\xspace}

\def\yago{Yago\xspace}
\def\nyt{NYT\xspace}

\DeclareCaptionType{copyrightbox}

\begin{document}
\maketitle


\begin{abstract}
We consider the problem of similarity search within 
a set of top-k lists under the Kendall's Tau distance function.
This distance describes how related two rankings
are in terms of concordantly and discordantly ordered
items. As top-k lists are usually very short
compared to the global domain of possible items to 
be ranked, creating an inverted index to look up
overlapping lists is possible but does not capture
tight enough the similarity measure. In this work,
we investigate locality sensitive hashing
schemes for the Kendall's Tau distance and evaluate
the proposed methods using two real-world datasets.
\end{abstract}
\section{Introduction}
\label{sec:intro}

Generating rankings is a well known methodology to order a set of elements 
to allow users or tools to immediately investigate the best performing elements,
according to the applied ranking criterion. This is frequently done,
in portals such as ranker.com that aim at crowdsourcing (subjective) entity rankings,
in data warehousing environments that rank business objects on objective/measurable criteria,
or given in form of tables on the Web. 
Rankings can serve ad-hoc information demands or give access
to deeper analytical insights.  Consider for instance
mining semantically similar Google-style keyword queries based on the 
query-result lists, or dating portals that let users create favorite lists
that are later-on used for match making. Such rankings are usually rather
short instead of exhaustively ranking the global domain or rank-able items.
This work emphasizes on Kendall's Tau as the distance measure for retrieving 
rankings in nearest neighbor (NN) 
search---for a user-given distance threshold and a ranking that serves
as the query. Although the Kendall's Tau distance is originally defined over
pairs of rankings that capture the same (full) domain of elements,
a definition of the generalized Kendall's Tau distance for top-k list is given by
Fagin et al. in~\cite{DBLP:journals/siamdm/FaginKS03}.
The authors in the same work also show that the 
generalized Kendall's Tau distance violates the triangle equality, hence, is not a metric,
so, using metric space index structures, like the M-tree~\cite{DBLP:conf/vldb/CiacciaPZ97}, 
is discarded immediately. Using the fact that at least one element
should be contained in both rankings in order to have a reasonable, minimum similarity,
one classical solution of NN search is to use inverted indices. Such indices are
very efficient in answering set-containment queries~\cite{DBLP:journals/vldb/HelmerM03}.
On the other hand, locality sensitive hashing (LSH) performs efficiently in approximate 
NN search for high-dimensional data.  In literature, there exist various hash function
families for different metric distances such as $l_1$, Euclidean,
or Hamming distance~\cite{DBLP:conf/compgeom/DatarIIM04,DBLP:conf/stoc/IndykM98}.
Although the Kendall's Tau  distance is a non-metric distance function, 
we observe a similarity between Kendall's Tau and the Hamming distance 
and also the Jaccard distance which encouraged 
us to work on LSH hash function families for the Kendall's Tau distance.

\subsection{Problem Statement and Setup}

We consider a set of rankings $\mathcal{T}$, where each $\tau_i \in \mathcal{T}$
has a domain $D_{\tau_i}$ of fixed size $k$. 
The global domain of items is then $D = \bigcup_{\tau_i \in \mathcal{T}} D_{\tau_i}$. 
We investigate the impact of various choices of $k$ on the query performance
in the experiments. For instance, we have the following input given:

\begin{center}
\begin{tabular}{|cl|}   \hline
\multicolumn{2}{|c|}{$\mathcal{T}$} \\
{\bf id} & {\bf ranking content}  \\ 
$\tau_1$ & $[2,5,4,3]$ \\ 
$\tau_2$ & $[1,4,7,5]$ \\ 
$\tau_3$ & $[0,8,7,5]$ \\  \hline
\end{tabular}
\end{center}

Rankings are represented as arrays or lists of items; where the left-most position, denotes the top ranked item.
The rank of an item $i$ in a ranking $\tau$ is given as $\tau(i)$.  

At query time, we are provided with a query ranking $q$, where $|D_q|=k$
and $D_q \subseteq D$, a distance threshold $\theta_d$, and distance
function $d$. Our objective is to find all rankings which belong to 
$\mathcal{T}$ and has distance less than or equals to 
$\theta_d$, i.e, 
$$\{\tau_i| \tau_i \in \mathcal{T} \wedge d(\tau_i, q)\leq \theta_d \}$$

As mentioned above, rankings can be interpreted as short sets
and we can build an inverted index over them,
to look up at query time those rankings 
that have at least one item overlap with the query's items.

Considering the above example, for a query ranking 
$q = [8, 1, 0, 6]$, ranking $\tau_1$ does not overlap at all with the query items, 
while $\tau_2$ and $\tau_3$ do overlap. The retrieval of overlapping candidates using an inverted
index is very efficient~\cite{DBLP:journals/vldb/HelmerM03}.
For the found candidates, the distance function is applied with respect to 
the query and the true results are returned.
Note that, we assume that the distance threshold $\theta_d$
is strictly smaller than the maximum possible distance (normalized, 1),
thus, in fact the inverted index can find all result rankings.

However,  Kendall's Tau is defined as the pairwise disagreement
between two permutation of a set, which suggests building an inverted 
index that is labeled (as keys) by pairs of items.
This, however, calls for investigating if looking up
at query time only a few pairs is sufficient.

\subsection{Contributions and Outline}

 Here, we have summed up the main contributions of this work.
\begin{itemize}
 \item We propose two different hash families for Kendall's Tau distance that facilitate LSH for NN search. 
 \item We compare the performance of those LSH schemes and traditional inverted indices
 by an experimental study using real-world rankings.
\end{itemize}

The paper is organized as follows. Section~\ref{sec:background} gives a brief
overview on the main principles of Kendall's Tau,  locality sensitive hashing,
and inverted indices. Section~\ref{sec:workingmodel} presents 
the use of a plain inverted index for query processing
and derives a distance bound for improved efficiency.
Section~\ref{sec:pairs} shows the consequences of interpreting
rankings as sets of pairs and motivates the derivation of LSH
schemes presented in Section~\ref{sec:hashingscheme}.
Section~\ref{sec:experiments} reports on the details of a conducted
experimental evaluation using two real-world datasets.
Section~\ref{sec:relatedwork} discusses related work.
Section~\ref{sec:conclusion} concludes the paper.
\section{Background}
\label{sec:background}

\subsection{Kendall's Tau on Top-k Lists}

Complete rankings are considered to be permutations  over a fixed 
domain $\mathcal{D}$. We follow the notation by Fagin et
al.~\cite{DBLP:journals/siamdm/FaginKS03} and references within. 
A permutation $\sigma$ is a bijection from the domain 
$\mathcal{D}= \mathcal{D}_\sigma$ onto the set $[n]=\{1, \ldots, n\}$.
For a permutation $\sigma$, the value $\sigma(i)$ is interpreted as 
the rank of element $i$. An element $i$ is said to be ahead of an
element $j$ in $\sigma$ if $\sigma(i)<\sigma(j)$.
The Kendall's Tau distance $K(\sigma_1, \sigma_2)$ measures how both 
rankings differ in terms of concordant and discordant pairs:
For a pair $(i,j)\in D\times D$ with $i\neq j$ we let
$\bar{K}_{i,j}(\sigma_1, \sigma_2) = 0$ if $i$ and $j$ are in the same
order  in $\sigma_1$ and $\sigma_2$
and $\bar{K}_{i,j}(\sigma_1, \sigma_2) = 1$ if they are in opposite order.
Then Kendall's Tau is  given as 
$K(\sigma_1, \sigma_2) = \sum_{i,j} \bar{K}_{i,j}(\sigma_1, \sigma_2)$.

Kendall's Tau is a metric, that is, it has the symmetry property,
i.e., $d(x,y)=d(y,x)$, is regular, i.e., $d(x,y)=0$ iff $x=y$,
and suffices the triangle inequality $d(x,z) \leq d(x,y)+d(y,z)$, 
for all $x,y,z$ in the domain.

In this work, we consider incomplete rankings, called top-k lists 
in~\cite{DBLP:journals/siamdm/FaginKS03}. Formally, a top-k list $\tau$ is
a bijection from $D_\tau$ onto $[k]$. The key point is that individual
top-k lists, say $\tau_1$ and $\tau_2$ do not necessarily share the same
domain, i.e., $D_{\tau_1} \neq D_{\tau_2}$. 
Fagin et al.~\cite{DBLP:journals/siamdm/FaginKS03} discuss how the
above two measures can be computed over top-k lists. None of the
discussed ways to compute Kendall's Tau over top-k lists is a metric.
 
This work considers incomplete rankings (lists) and applies the generalized
Kendall's Tau distance function defined by Fagin~\cite{DBLP:journals/siamdm/FaginKS03}.
Given two top-k lists $\tau_1$ and $\tau_2$ that correspond
to  two permutations  $\sigma_1$ and $\sigma_2$ on $D_{\tau_1}\cup D_{\tau_2}$,
the generalized Kendall's Tau distance with penalty zero,
denoted as $K^{(0)}(\tau_1, \tau_2)$ is defined as follows:
\begin{itemize}
\item If $i,j \in D_{\tau_1}\cap D_{\tau_2}$ and their order is the same in both list then $\bar{K}^{(0)}(\tau_1, \tau_2)=0$ otherwise $\bar{K}^{(0)}(\tau_1, \tau_2)=1$.
\item If $i,j \in D_{\tau_1}$ and $i$ or $j \in D_{\tau_2}$, let $i \in D_{\tau_2}$ and $\tau_1(i)<\tau_1(j)$ then  $\bar{K}^{(0)}(\tau_1, \tau_2)=0$ otherwise $\bar{K}^{(0)}(\tau_1, \tau_2)=1$.
\item If $i\in D_{\tau_1}$ and $j\in D_{\tau_2}$ or vice versa then $K^0(\tau_1, \tau_2)=1$.
\item If $i,j \in D_{\tau_1}$ and $i,j \notin D_{\tau_2}$ or vice versa $\bar{K}^{(0)}(\tau_1, \tau_2)=0$.
\end{itemize}

\subsection{Locality Sensitive Hashing}

Locality sensitive hashing addresses the problem of efficient approximate nearest 
neighbor (NN) search. The key idea is to 
map objects into buckets via hashing with the property that
similar objects have a higher chance to collide (in the same bucket) 
than dissimilar ones. A large amount of
research has been conducted on LSH in order to find fast and robust 
families of hash functions for different distance functions. 

Let us consider family $\mathcal{H}$ of hash functions that map a point
$p \in \mathbb{R}^d$ to some universe $U$. For threshold $r$ and approximation factor $c$, $\mathcal{H}$ is called $(r, cr, P_1, P_2)$-sensitive if 
for any two point $p, q \in \mathbb{R}^d$
\begin{itemize}
 \item if $p \in B(q,r)$ then $Pr_{\mathcal{H}}(h(p) = h(q)) \geq P_1$
 \item if $p \notin B(q, cr)$ then $Pr_{\mathcal{H}}(h(p) = h(q)) \leq P_2$
\end{itemize}

$B(q, r)$ represents a ball that is centered at $q$ and has radius $r$,
i.e., if a point $p$ is in $B(q,r)$ then its distance to  $q$ 
 is at most $r$. In order for LSH to be useful
for dissimilarity measure (where $c>1$) we should have $P_1 > P_2$ and $r < cr$.

Instead of employing only one hash function to determine
the bucket to put an object in, several $h_i$ out of
$\mathcal{H}$ are used to create a label 
$g(p) = (h_1(p), h_2(p), \ldots ,h_m(p))$.
Two objects $p_1$ and $p_2$ are put into the same
bucket if, obviously, $g(p_1)=g(p_2)$, hence, the more $h_i$ are used in $g$,
the fewer objects a bucket will contain. On the other hand,
if two objects are placed in the same bucket, the chance that they are
really similar is increased.

To counter the problem of suffering from low recall (i.e., the fraction
of results found) $l$ different 
hash tables are created  using $l$ different $g_j$ functions.
Then, for a query point $q$, if points $p$ and $q$ are both
 hashed to the same bucket
in {\it any} of the $l$ hash tables, $p$ is considered a potential
candidate. Finally, if $o \in B(q, r)$ then it is consider as near neighbor
of $q$, otherwise not.   

 \subsection{Inverted Index}
 \label{subsec:invindex}

Rankings can be interpreted as plain sets, ignoring the order among items.
One way to index sets of items is to create a mapping of items
 to the rankings in that the items are contained in. This resembles
the basic inverted index known from information retrieval and also 
used for querying set-valued attributes~\cite{DBLP:journals/vldb/HelmerM03}.

An inverted index consists of two components---a \emph{dictionary} $\mathcal{D}$ of \emph{objects} and the
corresponding \emph{posting lists} (aka. index list) that record for each object
information about its occurrences in the relation (cf., \cite{DBLP:journals/csur/ZobelM06} for an overview and implementation details).

For a given item, the inverted index is accessed and returns all rankings that contain the item.
\section{Inv. Index with Distance Bounds}
 \label{sec:workingmodel}

In this section, we  discuss how inverted 
indices can be used for computing Kendall's Tau and derive distance bounds
for improved performance. We use a basic inverted index on rankings,
shown in Table~\ref{table1}, as a baseline in the experimental evaluation.

\begin{table*} 
\begin{minipage}[b]{60mm}
 \begin{center}
 \begin{tabular}{|c|}
 \hline
 $7 \rightarrow \langle\tau_2\rangle, \langle \tau_3\rangle $ \\
 $5 \rightarrow \langle \tau_1\rangle, \langle \tau_2 \rangle, \langle \tau_3\rangle $ \\
  $4\rightarrow \langle\tau_1\rangle, \langle \tau_2\rangle$ \\
  $\ldots$\\
  \hline
 \end{tabular}
 \end{center}
 \caption{Basic inverted index}
 \label{table1}
\end{minipage}
\begin{minipage}[b]{60mm}
\begin{center}
\begin{tabular}{|c|}
 \hline
 $(4, 5) \rightarrow \langle\tau_1 \rangle, \langle \tau_2 \rangle $ \\
 $(5, 7) \rightarrow \langle \tau_2\rangle, \langle \tau_3  \rangle $ \\
  $(3, 4)\rightarrow \langle\tau_1\rangle $ \\
  $\ldots$\\
  \hline
 \end{tabular}
\end{center}
 \caption{Sorted pairwise index}
 \label{table2}
\end{minipage}
\begin{minipage}[b]{60mm}
\begin{center}
\begin{tabular}{|c|}
 \hline
  $(5, 4) \rightarrow \langle\tau_1 \rangle $ \\
 $(7, 5) \rightarrow \langle \tau_2\rangle, \langle \tau_3  \rangle $ \\
  $(4, 5)\rightarrow \langle\tau_2\rangle $ \\
  $\ldots$\\
  \hline
 \end{tabular}
\end{center}
 \caption{Unsorted pairwise index}
 \label{table3}
\end{minipage}
\end{table*}

Finding similar rankings for user given query ranking $q$ and a 
distance threshold $\theta_d$ follows a simple filter and validate pattern: 
\begin{itemize}
 \item The inverted index is looked up for each element in $D_q$ and
 a candidate set $\mathcal{C}$ of rankings is built by collecting all 
 distinct rankings seen in the accessed posting lists.
 \item For all such candidate rankings $\tau \in \mathcal{C}$,
 the distance function $K^{(0)}(\tau,q)$ is calculated and if 
  $K^{(0)}(\tau,q) \leq \theta_d$ then  $\tau$  is added to  result set $\mathcal{R}$.
\end{itemize}

Potentially very many of the candidate rankings in $\mathcal{C}$ are so called
{\it false positives}, i.e., rankings that are accessed but do not belong
to the final $\mathcal{R}$. Each such false positive causes an unnecessary
distance function call. 
Intuitively, final $\tau \in \mathcal{R}$ should be found in at least a certain 
number of posting lists, depending on distance threshold $\theta_d$,
and, in fact, we can derive such a criterion that removes
some of the  false positives but does not introduce {\it false
negatives}, i.e., missed results.

Assume $n$ elements are common between the query ranking $q$ and a specific ranking $\tau$. 
Then, the smallest possible $K^{(0)}(\tau,q)$ value is $(k-n)^2$,
considering all matched pairs of elements are in same order in
both the query  and the ranking and also, all missing elements of the query
and the ranking appear at the bottom of both lists. Note that
throughout the paper we use the non-normalized Kendall's Tau
distance, as in the definition of  $K^{(0)}(\tau,q)$ ; 
so in fact, $k^2$ is the maximum distance possible between two top-k.

We are interested in the least (minimum) number of common elements $\mu$ that are
required for a ranking to have the chance to be in $\mathcal{R}$. 
$\mu$ can be found from the ranking whose best score is $\theta_d$. 
Thus, solving the equation $(k-\mu)^2 = \theta_d$ for $\mu$, we get $\mu = k- \sqrt{\theta_d}$. 
Clearly, all rankings with ``overlap'' $n < \mu$ will have $K^{(0)}(\tau, q) > \theta_d$
and can be immediately ignored. As all final results
must appear in at least $\mu$ number of posting lists, we can avoid scanning 
$(\mu-1)$ number of elements from $D_q$.
Thus, in time of  building the candidate set, we can prune 
a significant amount of rankings just by looking into only $k-\mu+1$ posting lists.

\section{Rankings as Sets of Pairs}
\label{sec:pairs}

From the perspective of the definition of the Kendall's Tau distance, rankings 
can be viewed as a set of pair elements apart from set of ordered elements.
We consider two different representations of $\tau$ as set of pair elements. 
In general, indexing pairs is feasible as rankings are considered rather
short compared to the potentially large global domain. We will see below that
at query time not all pairs need to be used.

$\tau_u^p$ represents all pair of elements that occur in ranking $\tau$, defined as:
$$\tau_u^p = \{(i,j)| (i,j) \subseteq D_{\tau}\times D_{\tau} \wedge i<j\}$$ 
For example, $\tau_{1u}^p = \{(2,5), (2,4), (2,3), (4,5),(3,5) \ldots\}$.
The pair entries are sorted in lexicographic order for removing redundant indexing.
A pairwise index structure is proposed by mapping each pair $(i,j)\in \tau_u^p$ to a posting list that holds all 
rankings (ids) in which both the elements $i$ and $j$ occur.
For clarification, Table~\ref{table2} represents part of the index for the example rankings
given earlier.

Again, a simple filter and validate technique can be used for this index structure.
In this case, we look up the index for all pair $(i,j) \in q^p_u$. 
As we can calculate $\mu$ for specific $\theta_d$ and $k$, 
it is sufficient to look up the index for all the pairs that include any 
one of the $\mu$ elements. Hence, we need to compare at most $\sum_{i=1}^{k-\mu +1}(k-i)$ 
number of pairs from query and can prune potentially very many false-positive candidates.

A slightly different representation of a ranking $\tau$ is defined below, 
denoted as $\tau_s^p$, where each pair holds the information about the ranking order
between them.  

$$\tau_s^p = \{(i,j)| \tau(i) \leq \tau(j) \wedge i,j \in D_{\tau}\}$$

For example, $ \tau_{1s}^p = \{(2,5), (2,4), (2,3), (5,4), \ldots\}$. 
Based on this representation, a sorted-pairwise inverted index is used
to map  $(i,j) \in \tau_s^p$ to a posting list.
A posting list for $(i,j)$  holds all those rankings in which $i$ occurs before $j$.
For clarification, Table~\ref{table3} represents part of the index for the 
example given in the introduction.

A query is processed in this index in exactly the same way as it is
processed for the unsorted pair index. 

In practice, we can retrieve all result candidates by scanning much fewer pair
of elements than the bound we have established above. For instance,
in the experiments, in some cases, we are able to find more than 99\% of 
the result candidates by scanning only 1 pair from query. 

In the next section, we discuss the reason behind these characteristics by
relating the pairwise inverted index structures to locality sensitive hashing.
For this, we define hash functions and reason about their locality sensitivity.

\section{LSH Schemes for Kendall's Tau}
\label{sec:hashingscheme}

In this section, we investigate two hash families for LSH under
the Kendall's Tau distance. 

\subsection{Scheme 1}

We introduce the first hash family denoted as $\mathcal{H}_1$, which contains 
projection with respect to elements of the global domain $D$. 
$h_i \in \mathcal{H}_1$ where $i \in D$ is defined as 

\begin{equation} 
   h_{i}(\tau)= 
\begin{cases}
   1,& i\in \mathcal{\tau} \\
   0,& \mathrm{otherwise}
\end{cases}
\end{equation} 

For this scheme, we define a function family  $\mathcal{G}_1$ as

$$\mathcal{G}_1 = \{(h_i,h_j)| (h_i,h_j) \in  \mathcal{H}_1 \times \mathcal{H}_1\ \mathrm{and}\:i<j\}$$

That is, $g \in \mathcal{G}_1$ where $g = (h_i,h_j)$ projects
a ranking $\tau$ to $\{0,1\}^2$. In practice, we always project on
two elements that actually occur in query, i.e., look up the bucket label
$(1,1)$ for a specific $g$.
Clearly, this bucket for $g_1 = (h_i,h_j)$ is represented by key $(i,j)$ in the unsorted 
pairwise index. Now, looking up the index for $l$ number of pair
$(i,j) \in q_u^p$ means applying $l$ different hash functions $g \in \mathcal{G}_1$.
For different $l$, the query performance is compared in the experimental evaluation.

\subsubsection{Locality Sensitivity}
From Section~\ref{sec:workingmodel}, we can compute the
overlap $(\mu)$ that is required to have a chance to satisfy
distance threshold $\theta_d$.
 As $h_i \in \mathcal{H}_1$ maps 
$\tau,q \in \mathcal{T}$ to $\{0,1\}$ according to the presence of $i$ 
in $\tau$ and $q$, the probability $Pr[h(q)=h(\tau)]$ becomes 
the Jaccard similarity between them, which is $P_1 = \mu/(2k-\mu)$. 
We need at least a Jaccard similarity of $P_1$ between the query and the ranking,
which yields at most the Jaccard distance $(1-P_1)$, so, here, $r=(1-P_1)$.
When $\mu$ increases, i.e., the Jaccard distance decreases between rankings,
then $P_1$ increases (as denominator of $P_1$ decreases and numerator increases).
Then, for $cr$, we have $P_2=1-c(1-P_1)$. Thus, 
as long as the approximation factor $c$ is strictly larger than $1$,
 we get  $P_1>P_2$. Thus, the locality sensitive property holds for $\mathcal{H}_1$.

\subsection{Scheme 2}

Here, we use a hash function family $\mathcal{H}_2$ that contains all projection 
 based on all combination of pair elements, represented as %
$D_{\mathcal{P}}= \{(i,j)| (i,j) \in D\times D \text{ and }\ i<j\}$.

$$\mathcal{H}_2 = \{h_{i,j}| (i,j) \in D_{\mathcal{P}}\}$$

$K^{(0)}(\tau,q)$ is defined by the number of discordant pairs on the  domain $D_{\tau} \cup D_q$,
for a ranking $\tau$ and query ranking $q$, as described in Section~\ref{sec:background}.
 Here, we define hash functions $h_{ij} \in \mathcal{H}_2$ that project $\tau$ to $\{0,1\}$.

\begin{equation} 
   h_{ij}(\tau)= 
\begin{cases}
   1,& i,j \in \tau \text{ and }\ \tau(i)<\tau(j)  \\
   1,& i \in \tau \\
   0,& \mathrm{otherwise}
\end{cases}
\end{equation}

For this scheme, the hash function family $\mathcal{G}_2$ is defined by selecting 
any hash function over $\mathcal{H}_2$, i.e., 
$\mathcal{G}_2 = \{h_{ij}| h_{ij} \in \mathcal{H}_2\}$. 
For a  $g\in \mathcal{G}_2$, i.e., $g = \{h_{ij}\}$, the bucket labels 
`1' and `0' of $g$ are represented respectively by the key element $(i,j)$ 
and $(j,i)$ in the sorted pairwise index. Thus, 
looking up $(a,b) \in q_s^p$ in the sorted pairwise index is the same
as considering the bucket where $q$ is projected by a
$g = {h_{i,j}}$ with  $\{i,j\}=\{a,b\}$. Now, as above, 
we can say that doing so for $l$ number of pairs from $q_s^p$
 means applying $l$ different hash functions $g \in\mathcal{G}_2$ 
on the query ranking. The impact of $l$ is studied in the experimental
evaluation.

\subsubsection{Locality Sensitivity}

After projecting it on $D_{\mathcal{P}}$,
a ranking is represented as a string of $\{0,1\}$. 
For clarification, such representation for $\tau_1$ and $\tau_2$ under
hash family $\mathcal{H}_2$ is shown in Table~\ref{table4}.

\begin{table} 
\begin{center}
 \begin{tabular}{|c|c|c|c|c|}
 \hline
 $ - $ & $(2,5)$ & $(4,5)$ & $(3,4)$ & $\ldots $ \\
 \hline
 $ \tau_1$ & $1$ & $0$ & $0$ & $\ldots $ \\
 \hline
  $\tau_2 $ & $0$ & $1$ & $0$ & $\ldots $ \\
  \hline
 \end{tabular}
 \end{center}
 \caption{projection of ranking under $\mathcal{H}_2$}
 \label{table4}
\end{table}

Clearly, the hamming distance between such a representation of rankings
is directly related with Kendall's Tau distance between them.
Hence, the probability $Pr[h(q)=h(\tau)] $ is equal to the number of
projection on which $\tau$ and $q$ agree. 
As we consider incomplete, size $k$ rankings, the maximum number of
pair to investigate between two ranking is $k^2$. Here, $r=\theta_d$ and we obtain $P_1 = 1 - \theta_d /k^2 $. 
If the distance between rankings is smaller 
than $\theta_d$ then $P_1$ increases, i.e., if rankings are more similar
then probability to project those ranking into same bucket is more. 
Also, as long as $c>1$, we get $P_1>P_2$ and the property of locality sensitive hashing holds for $\mathcal{H}_2$.

\subsection{Comparisons of the Two Schemes}
 
The two presented schemes are compared in this section with respect
to the probability of projecting similar items to the same bucket. 

In general, as the function family $\mathcal{G}$ is created 
by concatenating $m$ number of hash function $h$ and 
$l$ different $g\in \mathcal{G}$ are applied, 
the probability of becoming a candidate pair is $(1-(1-P_1^m)^l)$.

For the first scheme, $\mathcal{G}_1$ is created by concatenating two
hash function from $\mathcal{H}_1$, so $m=2$. With $l=1$, the probability of
becoming a candidate is $f_1:= 1-(1-(\mu/(2k-\mu))^2)$.
Using the  $\mu$ as given in Section~\ref{sec:workingmodel}
and simplifying this, 
$f_1 = (k- \sqrt{\theta_d})^2/(k+ \sqrt{\theta_d})^2$.

For the second scheme, we know $m =1$. Considering $l=1$, the
 probability of becoming a candidate is $f_2 := (1-(1-(1 - \theta_d /k^2)))$ 
 which is simplified to $f_2 = 1 - \theta_d /k^2$ and
 $f_1/f_2 = k^2(k-\sqrt{\theta_d})/(k+\sqrt{\theta_d})^3 \leq 1$.
Hence, $f_1 \leq f_2$.
This is also reflected in experiments below.

\section{Experiments}
\label{sec:experiments}
We have implemented the index structures as described above in Java 1.7
and conducted the experiments using an Intel i5-3320M CPU @ 2.60GHz 
Ubuntu Linux machine (kernel 3.8.0-29) with 8GB memory. The index structures are kept entirely in memory.

To test the querying performance in terms of query response time (wallclock time),
number of retrieved candidates, and recall (fraction of
results found), we use two different datasets.
\begin{figure*}[!t]
    \centering

    \subfigure[$k=10$] {\label{fig:yago_10_kt}\includegraphics[width=0.22\textwidth]{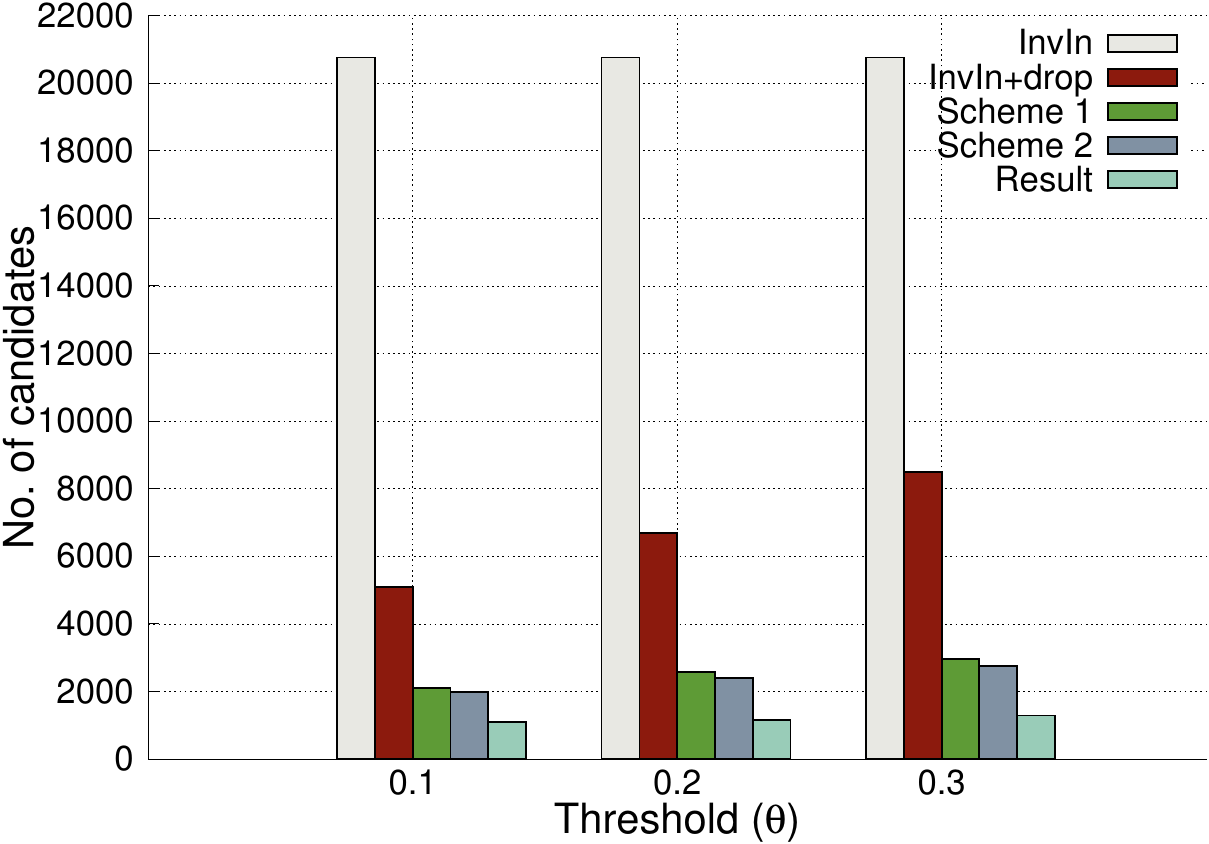}}
        \subfigure[$k=10$] {\label{fig:yago_10}\includegraphics[width=0.22\textwidth]{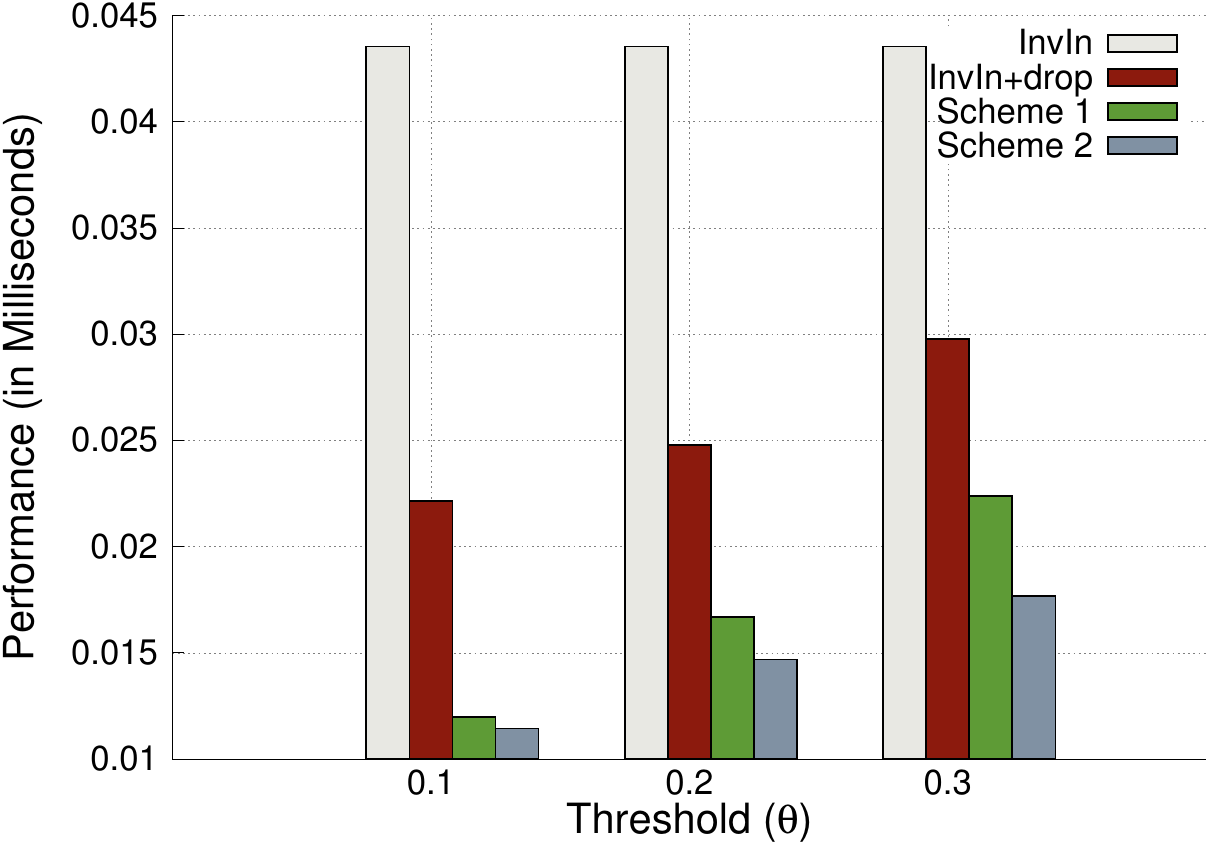}}
    \subfigure[$k=20$] {\label{fig:yago_20_kt}\includegraphics[width=0.22\textwidth]{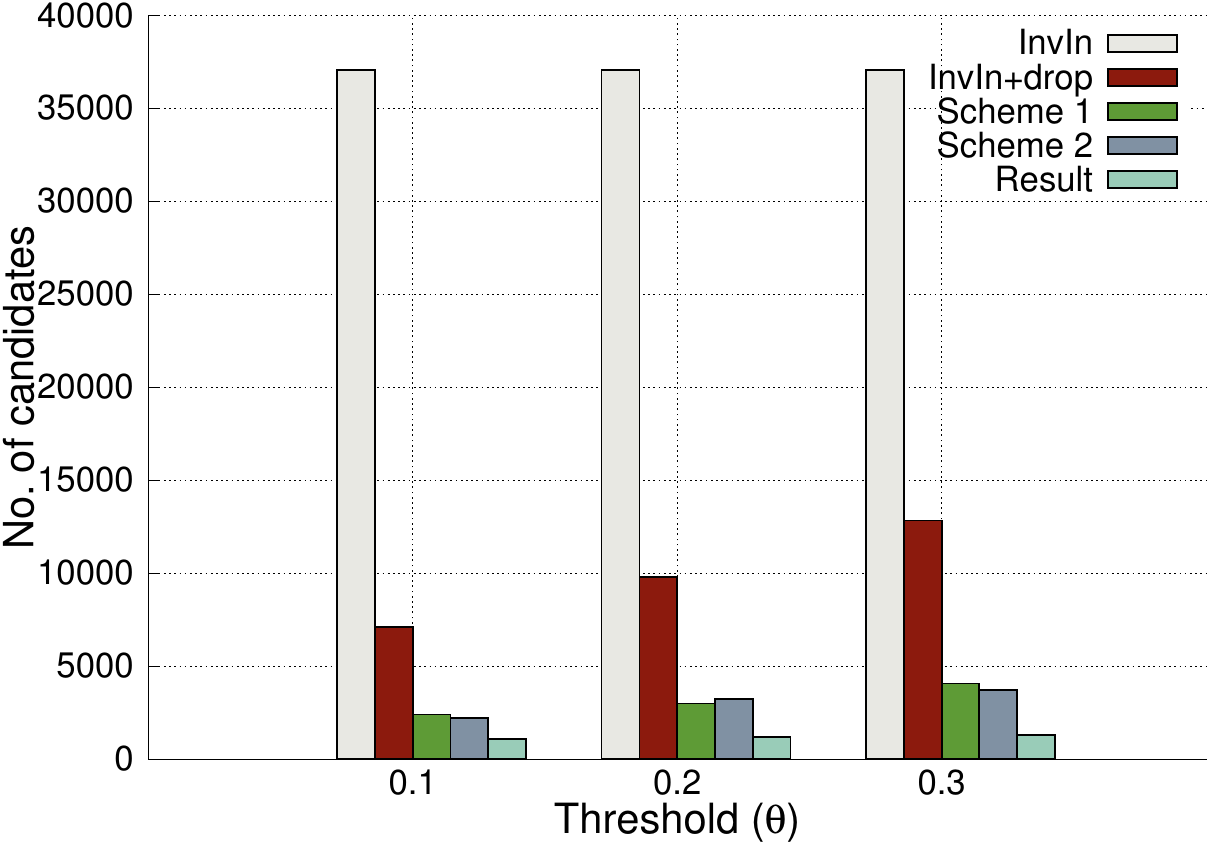}}
        \subfigure[$k=20$] {\label{fig:yago_20}\includegraphics[width=0.22\textwidth]{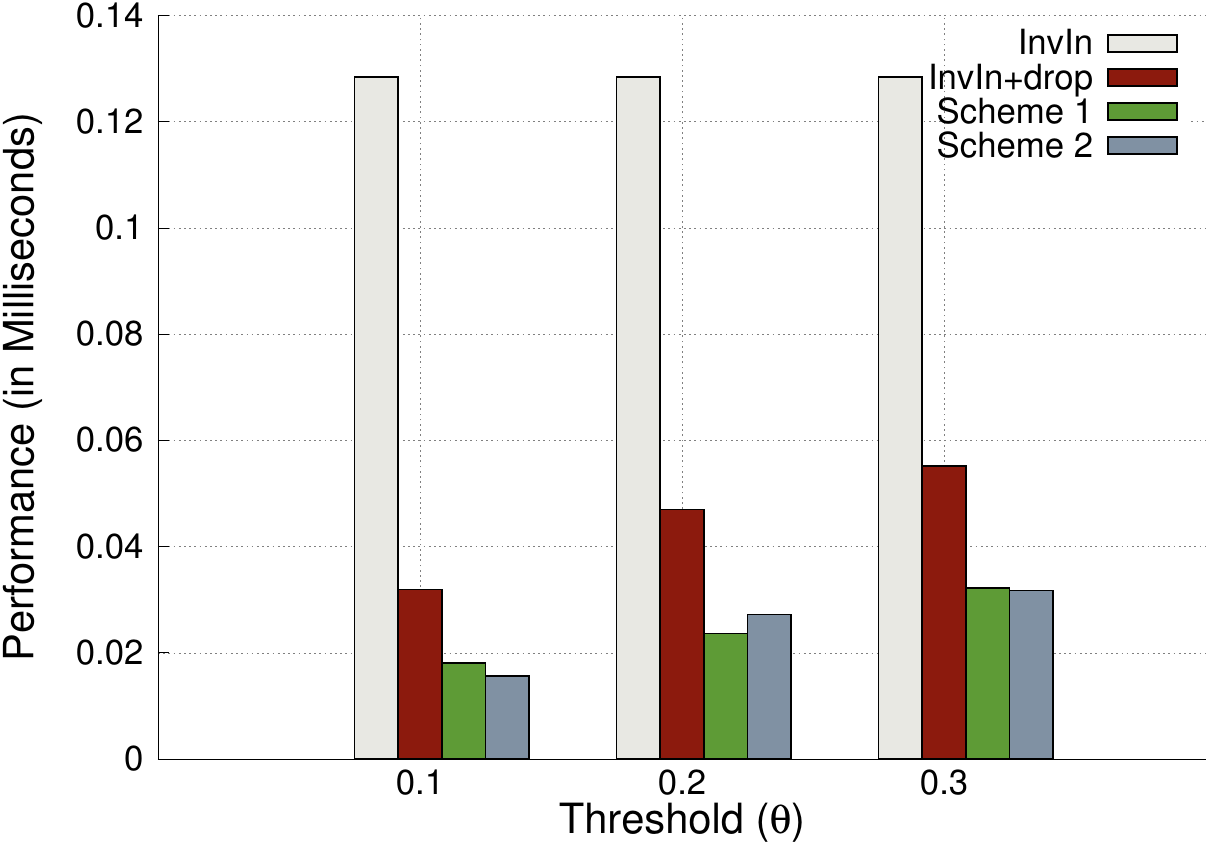}}

    \caption{Comparative study of query processing for varying $\theta$ (\yago).}
    \label{fig:yago}
\end{figure*}

\begin{figure*}[!t]
    \centering

    \subfigure[$k=10$] {\label{fig:nyt_10_kt}\includegraphics[width=0.22\textwidth]{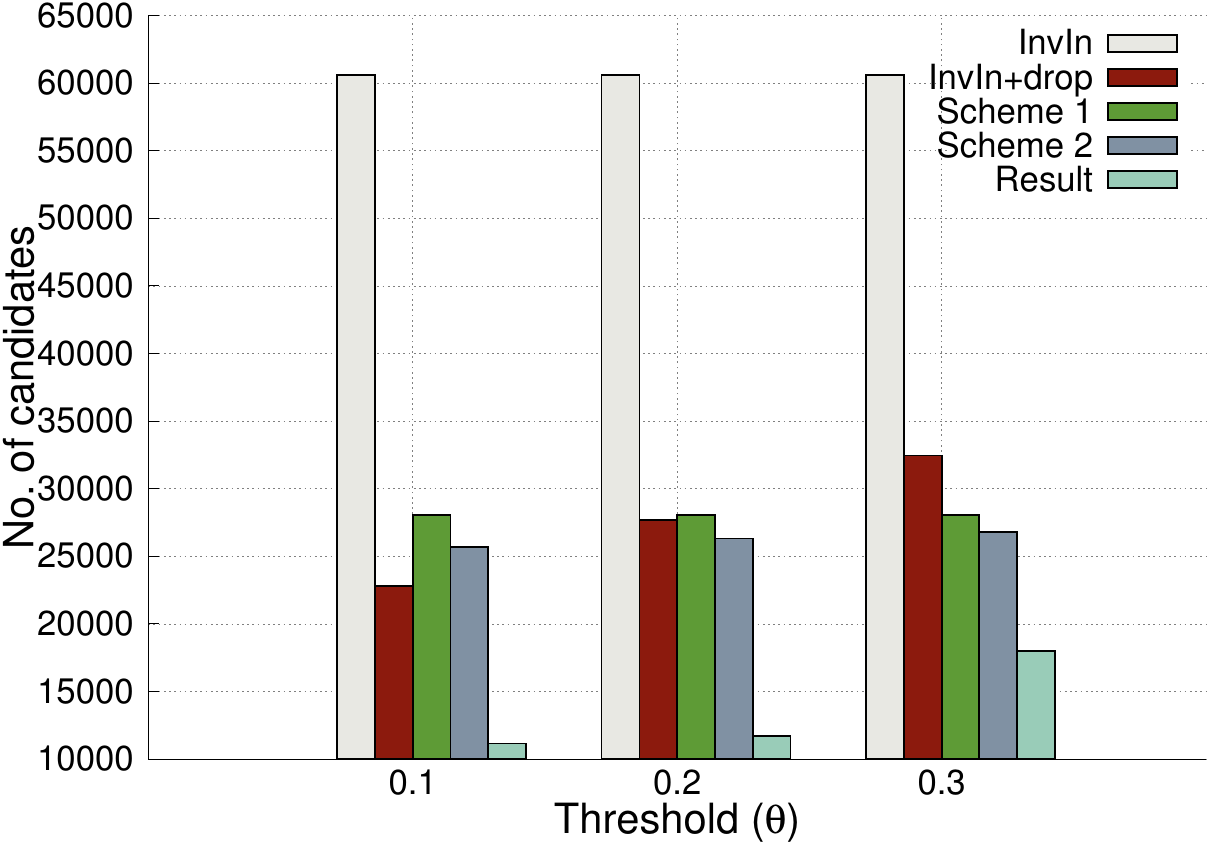}}
        \subfigure[$k=10$] {\label{fig:nyt_10}\includegraphics[width=0.22\textwidth]{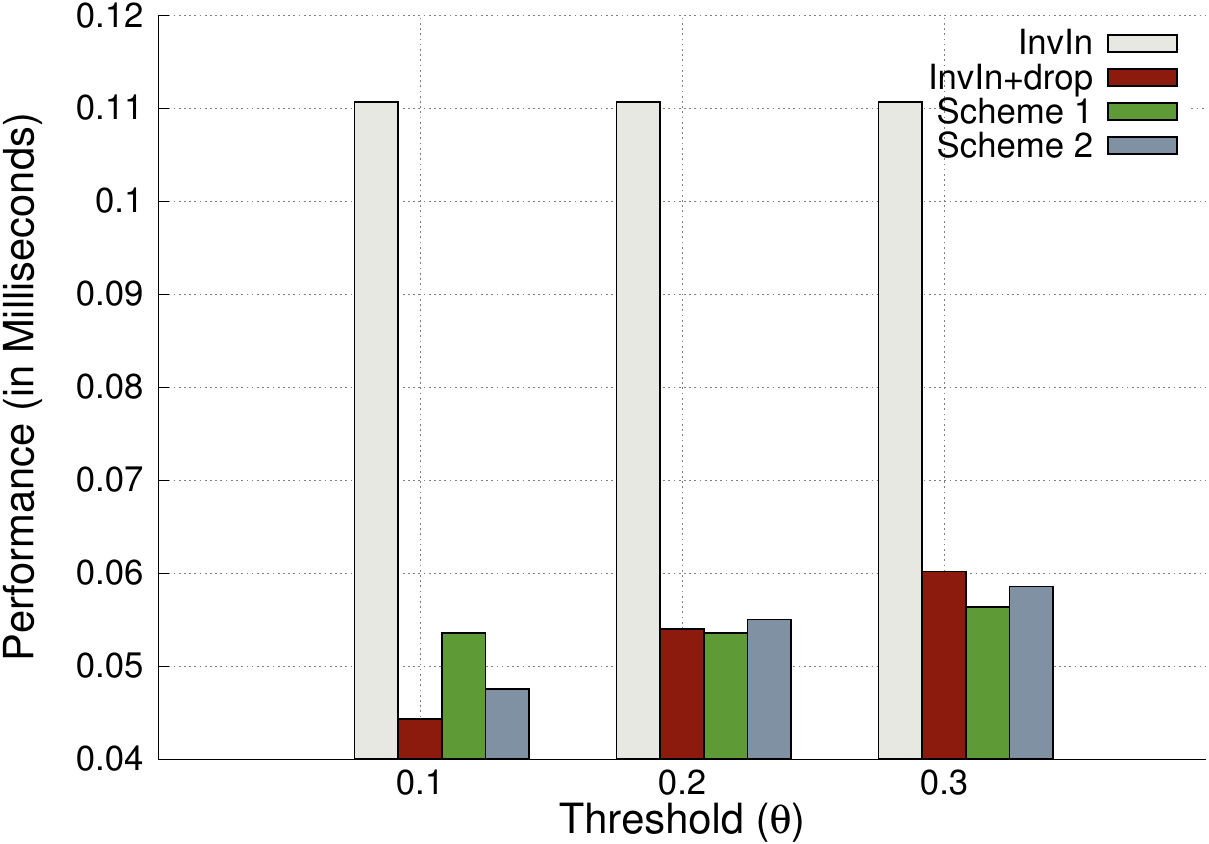}}
    \subfigure[$k=20$] {\label{fig:nyt_20_kt}\includegraphics[width=0.22\textwidth]{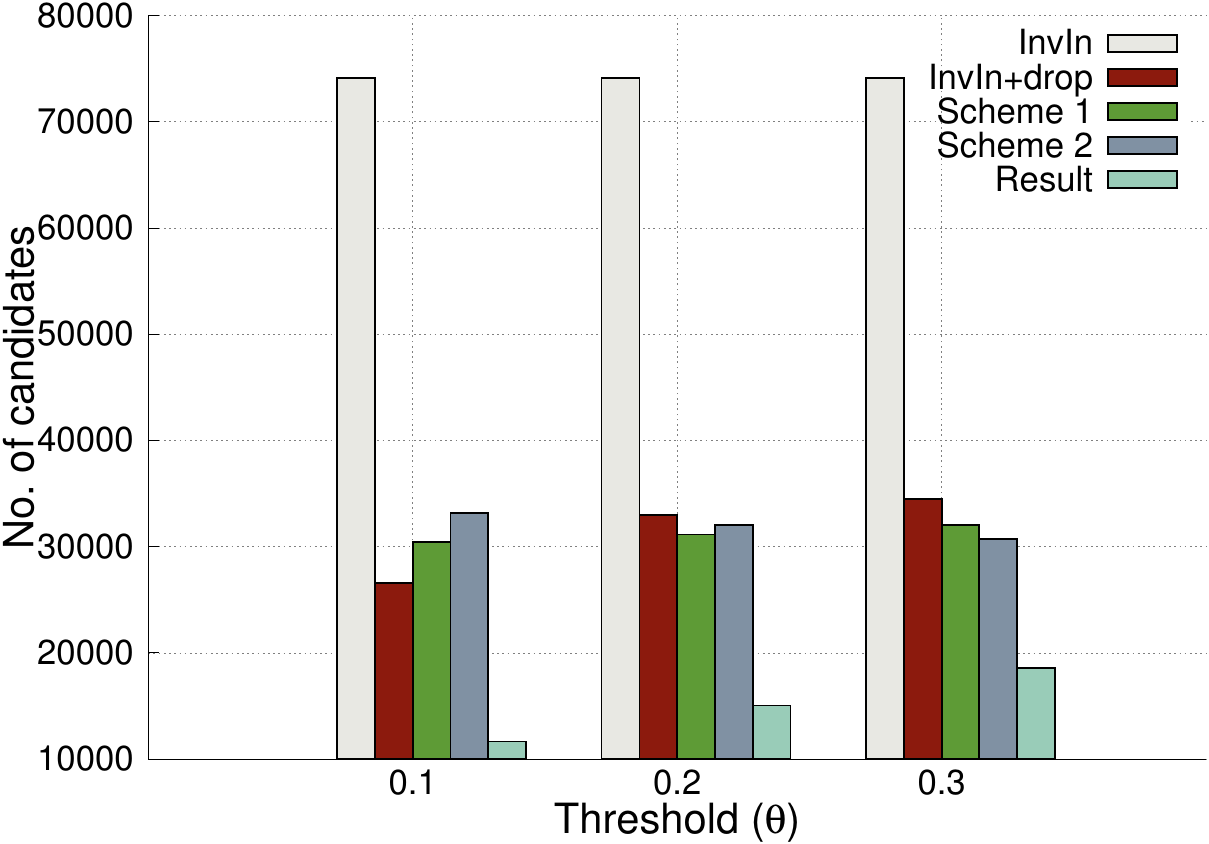}}
        \subfigure[$k=20$] {\label{fig:nyt_20}\includegraphics[width=0.22\textwidth]{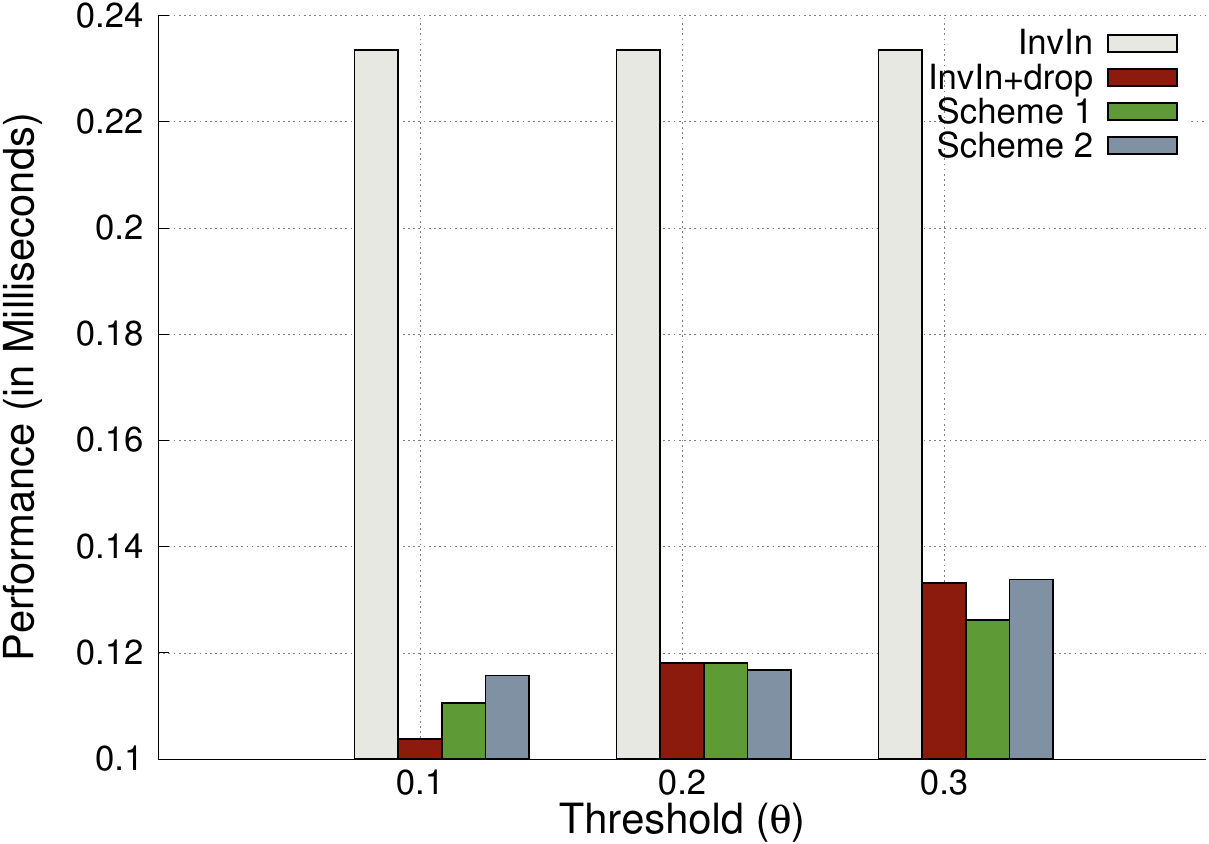}}

    \caption{Comparative study of query processing for varying $\theta$ (\nyt).}
    \label{fig:nyt}
\end{figure*} 

\begin{table*}[!t]
\centering
\begin{tabular}{|c|cccc|cccc|cccc|} 
\hline
& \multicolumn{4}{|c|}{\bf $\theta$ =0.1} &  \multicolumn{4}{|c|}{\bf $\theta$ =0.2} &  \multicolumn{4}{|c|}{\bf $\theta$ =0.3} \\ 
& l=1 & l=3 & l=6 & l=10 & l=1 & l=3 & l=6 & l=10 & l=1 & l=3 & l=6 & l=10 \\ \hline
Scheme 1 for \nyt  & 100 & 100 & 100 & 100 & 99.9 & 100 & 100 & 100 & 99.8 & 99.9 & 100 & 100 \\ \hline
Scheme 2 for \nyt & 99.7 & 100 & 100 & 100 & 98.9 & 99.8 & 100 & 100 & 97.9 & 99.6 & 99.8 & 100 \\ \hline 
Scheme 1 for \yago  & 98.9 & 100 & 100 & 100 & 97.1 & 99.5 & 100 & 100 & 92.1 & 97.9 & 99.9 & 100 \\ \hline
Scheme 2 for \yago & 98.7 & 100 & 100 & 100 & 96.6 & 99.3 & 100 & 100 & 91.3 & 97.3 & 99.7 & 100 \\ \hline 
\end{tabular}$\,$\\
\caption{Comparison of achieved recall in percent for $k=10$}
\label{table5}
\end{table*}

\begin{table*}[!t]
\centering
\begin{tabular}{|c|cccc|cccc|ccccc|} 
\hline
& \multicolumn{4}{|c|}{\bf $\theta$ =0.1} &  \multicolumn{4}{|c|}{\bf $\theta$ =0.2} &  \multicolumn{5}{|c|}{\bf $\theta$ =0.3} \\ 
& l=1 & l=3 & l=6 & l=10 & l=1 & l=3 & l=6 & l=10 & l=1 & l=3 & l=6 & l=10 & l=15 \\ \hline
Scheme 1 for \nyt  & 100 & 100 & 100 & 100 & 99.9 & 100 & 100 & 100 & 99.2 & 99.9 & 100 & 100 &100 \\ \hline
Scheme 2 for \nyt & 99.7 & 100 & 100 & 100 & 98.6 & 99.5 & 99.9 & 100 & 97.5 & 99.2 & 99.8 & 100 & 100 \\ \hline 
Scheme 1 for \yago  & 99.1 & 100 & 100 & 100 & 96.4 & 98.8 & 99.9 & 100 & 92.0 & 96.8 & 99.2 & 99.6 & 99.9 \\ \hline
Scheme 2 for \yago & 99.0 & 100 & 100 & 100 & 95.6 & 98.4 & 99.5 & 99.8 & 90.8 & 96.3 & 98.9 & 99.5 & 99.9 \\ \hline 
\end{tabular}$\,$\\
\caption{Comparison of achieved recall in percent for $k=20$}
\label{table6}
\end{table*}

{\bf \yago Entity Rankings}:
This dataset contains 25,000 top-k rankings which has been 
mined from the \yago knowledge base, as described in~\cite{DBLP:conf/cikm/IlievaMS13}. 

{\bf \nyt:} This dataset contains 1 million keyword query
 that are randomly selected out from a published query log of a 
large US Internet provider, against the New York 
Times archive~\cite{link:nytimescorpus} using a standard 
scoring model from the information retrieval literature. 

The \yago dataset holds real world entities where each entity occur in few rankings 
while the NYT dataset holds many popular documents that appear 
in many query result rankings.

The following approaches are compared to each other:

\begin{itemize}
\item The filter and validate technique on the simple inverted index denoted as {\bf InvIn}.
\item The InvIn technique on the simple inverted index combined with dropping some posting
lists from consideration using the distance bound given in Section~\ref{sec:workingmodel}, denoted as {\bf InvIn+drop}.
\item The presented LSH scheme 1, i.e., the unsorted pairwise index, denoted as {\bf Scheme 1}.
\item The presented LSH scheme 2, i.e., the sorted pairwise index, denoted as {\bf Scheme 2}.
\end{itemize} 

For both LSH schemes, unless $l$ is explicitly stated, $l$ is tuned such that 100\% recall are reached.
Runtime performance is measured in terms of average runtime for $1000$ 
queries for varying the {\it normalized} distance threshold $\theta$ (given by $\theta_d=k^2\times\theta$). 

We see in Figure~\ref{fig:yago} for the \yago dataset
that using the pairwise indices much less candidates are retrieved
than for simple inverted indices. Since recall is tuned to 100\% this means
that much fewer false positives are evaluated. This happens 
as, according to the LSH technique, true positive candidates are more likely
to be hashed into the same bucket. This is consistent through
datasets and parameters except for $\theta=0.1$ (larger difference) and $\theta=0.2$ (almost exactly
the same performance) for $k=10$ for the \nyt dataset (cf., Figure~\ref{fig:nyt})
where InvIn+drop performs best. For the plots showing the number of 
retrieved candidates, we put the actual number of results to mark the lower bound.

These characteristics are also reflected in the runtime performance
for the \yago dataset, but varies for the \nyt dataset. For the latter,
in some cases (particularly for $\theta=0.1$)  both LSH schemes show
inferior performance compared to the InvIn+drop.

We  also see that the sorted pairwise index (Scheme 2)
consistently retrieves fewer candidates than the unsorted pairwise index (Scheme 1).
 This property reflects that for different $l$ values, probability of retrieving candidates
 that belong to $\mathcal{R}$ in Scheme 2 is higher than Scheme 1,
which has been theoretically shown in Section~\ref{sec:hashingscheme} for $l=1$.
Thus, in other way, Scheme 2 is less likely to find a 
true positive result than Scheme 1 for same $l$, which also reflects in  
Table~\ref{table5} and Table~\ref{table6}.

In addition, comparing the columns of the table, 
we see that the recall increases as $l$ increases; which is in line with the LSH theory. 
We also understand the characteristics of the datasets by analyzing the recall.
For both schemes, with the same threshold $\tau$ and value of $l$, the recall for the \nyt dataset 
is always larger or equal to the recall for the \yago dataset. 
This reflects that elements of the \nyt dataset are featured more skewed than 
in the \yago dataset.

\section{Related Work}
\label{sec:relatedwork}

There is an ample work on computing relatedness between ranked lists of items,
such as to mine correlations or anti-correlations between lists ranked
by different attributes. Arguably, the two most prominent similarity measures are Kendall's tau and
Spearman's Footrule. Fagin et al.~\cite{DBLP:journals/siamdm/FaginKS03} 
study comparing incomplete top-$k$ lists,  that is, lists capturing a subset of a global set of items, rendering the lists incomplete in nature.
In the scenarios
motivating our work, like similarity search favorite/preference rankings,
lists are naturally incomplete, capturing, e.g., only the top-10 movies of all times.
In this work, we focus on the computation of Kendall's Tau distance.

 Helmer and Moerkotte \cite{DBLP:journals/vldb/HelmerM03} present a study on indexing set-valued attributes
as they appear for instance in object-oriented data\-bases.  Retrieval is done based on the query's items;
the result is a set of candidate rankings, for which the distance function can be computed. 
For metric spaces, data-agnostic structures for indexing objects are known, like the M-tree by Ciaccia et al.~\cite{DBLP:conf/vldb/CiacciaPZ97,DBLP:journals/vldb/ZezulaSAR98}; but Kendall's Tau over incomplete list is not a metric.

Wang et al.~\cite{DBLP:conf/sigmod/WangLF12} propose an adaptive framework for similarity joins
and search over set-valued attributes, based on prefix filtering.
This framework can be applied in the filter and validate technique on the \naive inverted index discussed in this work. 
The proposed LSH schemes are related to the concept of prefix filtering with parameter 2;
a detailed investigation of this is part of our future work.
The key idea behind Locality Sensitive Hashing (LSH)~\cite{DBLP:conf/focs/AndoniI06, DBLP:conf/compgeom/DatarIIM04, DBLP:conf/vldb/GionisIM99} is the usage of locality preserving hash functions 
that map, with high probability, close objects to the same hash value (i.e., hash bucket). 
Different parameters of locality preserving functions together with the number of hash function used, render LSH a parametric approach. Studies concerning LSH parameter tuning \cite{DBLP:conf/cikm/DongWJCL08,
DBLP:conf/www/BawaCG05} have been performed providing an insight into LSH parameter tuning for optimal performance.
LSH can be extended for non-metric distance using reference object has explained in~\cite{DBLP:conf/icde/AthitsosPPK08}.

Diaz et al.~\cite{DBLP:conf/sigir/DiazMA10} consider matchmaking between users in a dating portal.
The attributes considered are scalar (e.g., age, weight, and height) or categorical (e.g., married, smoking, and education)
and focus is put on feature selection and learning for effective match making.
 Work on rank aggregation~\cite{DBLP:conf/www/DworkKNS01,DBLP:conf/alenex/SchalekampZ09}
 aims at synthesizing a representative ranking that minimizes  the distances to the given rankings,
 for a given input set of rankings.

\section{Conclusion}
\label{sec:conclusion}

In this paper, we proposed two different hash function families for 
querying for similar top-k lists with respect to the generalized Kendall's Tau distance. 
From the experimental results, we have concluded that the performance of
query processing using LSH scheme outperforms the original inverted
index for datasets where the entities  of a domain are ``uniformly ''
distributed in rankings, whereas the performance of the LSH schemes
is similar or sometimes inferior  on datasets where some popular entities appear 
in large amount of rankings. We also studied the differences
between the two proposed hashing schemes both theoretical and experimentally.
Further, we would like to investigate indexing beyond
pairs (i.e., using triplets and more) of ranked items
and to harness the derived distance bounds and learned influences
of $l$ on pruning the index size.


\bibliographystyle{abbrv}


\end{document}